\documentclass[journal=jacsat,manuscript=article]{achemso}
\usepackage{amsfonts,amsthm,amsmath,amssymb,bm}
\usepackage{multirow,tabularx,diagbox,slashbox}
\usepackage{url,hyperref}
\usepackage{graphicx}
\usepackage{color}
\author{Fanlong Meng}
\author{Robyn H. Pritchard}
\author{Eugene M. Terentjev}
\email{emt1000@cam.ac.uk}
\affiliation{Cavendish Laboratory, University of Cambridge, JJ Thomson Avenue, Cambridge CB3 0HE, U.K.}

\title{Stress relaxation, dynamics and plasticity of transient polymer networks}

\date{\today}

\begin{document}

\begin{abstract}
\noindent
We propose a theoretical framework for dealing with a transient polymer network undergoing small deformations, based on the rate of breaking and re-forming of network crosslinks and the evolving elastic reference state. In this framework, the characteristics of the deformed transient network at microscopic and macroscopic scales are naturally unified.
Microscopically, the breakage rate of the crosslinks is affected by the local force acting on the chain.
Macroscopically, we use the classical continuum model for rubber elasticity to describe the structure of the deformation energy, whose reference state is defined dynamically according to when crosslinks are broken and formed.
With this, the constitutive relation can be obtained. We study three applications of the theory in uniaxial stretching geometry: for the stress relaxation after an instantaneous step strain is imposed, for the stress overshoot and subsequent decay in the plastic regime when a strain ramp is applied, and for the cycle of stretching and release. We compare the model predictions with experimental data on stress relaxation and stress overshoot in physically bonded thermoplastic elastomers and in vitrimer networks.
\end{abstract}

\section{Introduction}

Transient networks, also called physical gels, play an important role in technology and in biological systems \cite{ross-murphy}. The unique ability to re-shape solid in an arbitrary way by plastic deformation at a higher temperature, returning back to a fully rubber-elastic state at lower temperatures without any permanent degradation, including self-healing of mechanical damage, is what makes this class of soft materials so attractive in a variety of biological substitutes and functional material applications.
In all cases there is some physical (non-covalent) bonding that holds such a network together; there are many examples of hydrogen or ionic bonding \cite{kremer2009,Noro2009}, and local  hydrophobic interactions \cite{Suzuki2012,Serero2000}, as well as effective crosslinking by semi-crystalline or amorphous phase-separated micelles \cite{winter1991,kremer2009,stadler1996}. Biological networks are often bonded by transient protein-protein interaction~\cite{Perkins2010, Stelzl2005}, or by filament-membrane interaction~\cite{Zimmermann2012, Bornschlogl2013}. The interest in elastic properties of transient networks with breakable crosslinks dates back to the early work of Thomas \cite{thomas1966}  and Flory \cite{flory1960}
which, at that time, mostly concentrated on hydrogen bonding crosslinks. Later much attention was given to thermoplastic elastomers of block-copolymers \cite{bates1990,Hotta2002,Chassenieux2011}. In all of the mentioned cases, physically bonded crosslinks break under stress and at elevated temperature.  Very recently, a new class of transient network was developed, and given the name `vitrimer', where the covalent bonds holding the polymer chains in the network can be re-arranged by transesterification reaction \cite{Montarnal,Capelot,Guan1012} or a catalyst-free transamination of vinylogous urethanes \cite{Denissen}. In these systems, the shape of the network can be re-molded at a sufficiently high temperature, yet the number of covalent crosslinks remains the same at all times.

\begin{figure}[t]
\centering
\includegraphics[width=0.55\columnwidth]{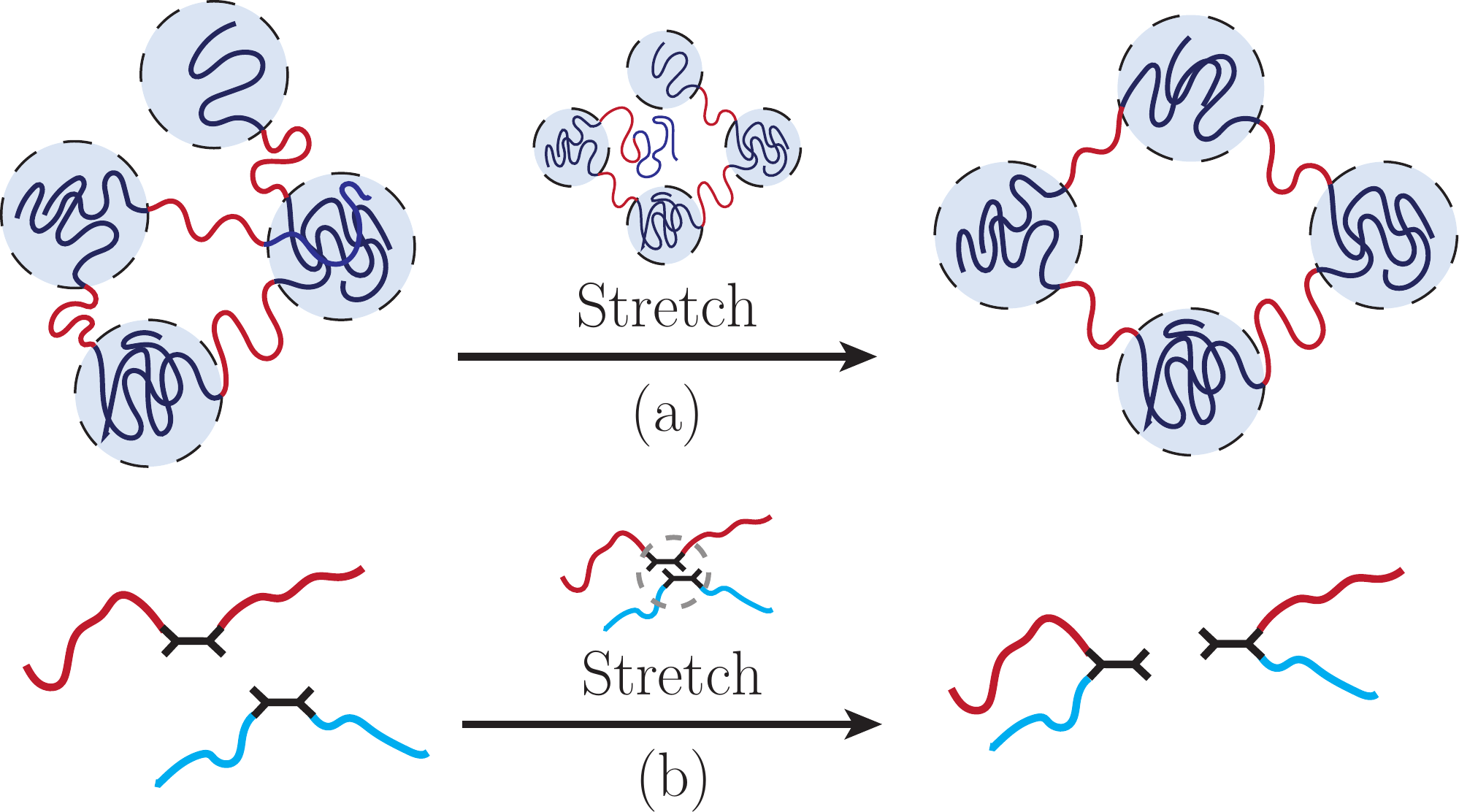}
\caption{Network rearrangements under stretch in: (a) physically crosslinked thermoplastic elastomer, and (b) covalently bonded vitrimer network re-configuring itself by transesterification.  } \label{phenomena}
\end{figure}

Figure~\ref{phenomena}(a) illustrates a way of effective network crosslinking via aggregates of chain segment, which could be in a crystalline, glass, or just rigid hydrogen-bonded arrangement. Figure~\ref{phenomena}(b) illustrates the topology of chain re-connection due to reversible covalent bonding such as transesterification, or transamination. Although the chemical nature of polymers involved, and the physical nature of crosslinks are very different, the common feature of all these materials is that they all have crosslinks that can be broken by force and spontaneously re-formed, usually after chain relaxation in a non-force-bearing configuration.

Theoretically, understanding the mechanics and relaxation in transient networks has been a long-standing project.
Microscopically, Green and Tobolsky~\cite{Green1946} introduced breakage and re-making of the crosslinks
when handling relaxation in polymeric networks, which was further developed by Fricker~\cite{Fricker1973} and Baxandall and Edwards~\cite{Baxandall1988}. Following this line of research, Tanaka and Edwards have put together a consistent framework of treating the crosslink dynamics under external force~\cite{Tanaka1992, Tanaka1992c}.
Separately, Rouse dynamics and reptation were used for studying the dynamics of a transient network by Leibler \emph{et al.}~\cite{Leibler1991}, later developed by Rubinstein and Semenov ~\cite{Rubinstein2001, Semenov2002}.

Macroscopically, in a series of papers, Drozdov \emph{et al.}~\cite{Drozdov1999, Drozdov2006} proposed constitutive models for various systems involving transient networks, by analyzing the macroscopic deformation energy.
In this approach one simply assumes appropriate expressions for the crosslink breakage and the re-forming rates
as a function of energy density with fitting parameters.
Similar ideas were successfully applied to deal with dual networks by Long and Hui \emph{et al.}~\cite{Long2014, Hui2012}, where the system consists of interpenetrating permanent and transient networks.
For simulations, Langevin dynamics~\cite{Noro2009, VandenBrule1995, Linder2011}, Monte Carlo~\cite{Groot1994, Groot1996} and molecular dynamics simulations~\cite{Khalatur1999, Khalatur1998, Hoy2009}
were applied to study the rheological behavior of a transient network.
It is usually simple to get a constitutive relation, if given the continuum/macroscopic energy form of the system.
If the microscopic details can be naturally incorporated into such a macroscopic picture,
then the theory can become portable and easy to be modified to meet customized conditions.

As it is known, the classical continuum model for rubber elasticity, sometimes called `neo-Hookean model',
can be obtained by statistically treating polymers as Gaussian chains~\cite{Treloar}.
In this work, we will follow Tanaka-Edwards method~\cite{Tanaka1992}
by explicitly applying the classical continuum model to describe the energy of the system,
instead of using complex statistical calculations. Specifically,
we can obtain the rate of the chains to break from crosslinks,
together with their re-crosslinking rate, by describing polymer chains as Gaussian (which is consistent with the level of approximation used in the neo-Hookean model).
By incorporating these molecular details into the time evolution of the macroscopic transient network structure,
we obtain the deformation energy of the system and then the constitutive relations under arbitrary geometry of strain. We then focus on the uniaxial stretching as an example (one of the most common geometries for study of dynamics and relaxation in experiment), and derive expressions for stress relaxation, ramp deformation and self-healing of the network in a cycle of deformation. In most cases we also carry out matching experiments on the SIS (styrene-isoprene-styrene) telechelic copolymer network physically crosslinked by glassy micelles of polystyrene~\cite{Hotta2002}, and on the classical transesterifying vitrimers of Leibler \textit{et al.}~\cite{Montarnal} Although this has never been studied in detail, one can assume that the rate of spontaneous re-crosslinking of broken-out chains is slow in SIS (where the chain end diffusion towards a new micelle needs to occur) and fast in vitrimers where the two chains simple re-connect in the same location. This comparison, which we can explicitly see in the analytical theoretical expressions, was the motivation for this choice. We find a good agreement with experiments, and discuss this and the implications at the end of the paper.

\section{The model}
In this section, we first describe the microscopic picture of rates of breakage and re-forming of crosslinks in a transient network under tension. We then derive the macroscopic elastic energy of the system, together with the general constitutive stress-strain relation, where the microscopic details of the crosslink dynamics are incorporated.

\subsection{Breaking and re-forming of a crosslink}

We shall work under a natural assumption that the crosslink is held together in a potential energy well with a characteristic energy barrier to overcome,  $W_{\mathrm{b}}$. The equilibrium Kramers rate of breakage of such a system is given by the thermally activated law
\begin{eqnarray}\label{rate11}
 \beta &=& \omega_{0}e^{-(W_{\mathrm{b}}-fb)/k_{\mathrm{B}}T},
\end{eqnarray}
where $\omega_{0}$ is the natural frequency of thermal vibration of the reactive group in the isolated state. The work by an external force $\bm{f}$ acting on the chain connected to this crosslink is obtained by assuming that a displacement of one monomer length, $b$, is enough to pass the confinement barrier.
For a Gaussian chain (a valid approximation in a polymer melt due to screening of self-interactions), the force acting on the chain is simple: $\bm{f}=3k_{\mathrm{B}}T\bm{r}/N_{s}b^{2}$,
where $\bm{r}$ is the end-to-end vector of the chain,
and $N_{s}$ is the number of the segments constituting a chain that connects the crosslinks.
Alternatively, the acting force can be obtained from the stress tensor,
which will be illustrated later.

Equation~\eqref{rate11} can also be arranged in the form that separates the exponential factor containing the applied force, and converts this force into the end-to-end distance of polymer strand connecting two crosslinks:
$ \beta = \beta_{0}e^{\kappa r }$,
where the parameter $\kappa = 3/N_{s}b$, and $\beta_0$ is the spontaneous breaking rate determined by the barrier $W_b$. The average end-to-end distance $\langle \bm{r} \rangle$ of a deformed network changes with imposed deformation $\mathbf{E}$, following the affine expression $\langle \bm{r} \rangle = \langle \mathbf{E} \cdot \bm{r}_0 \rangle$ with an appropriate orientational averaging, resulting in the dependence of the breaking rate on deformation.   When both breaking and re-forming of crosslinks takes place and the deformation is dynamic, $\mathbf{E} = \mathbf{E}(t)$, the breakage rate $\beta(t,t_{0})$ is a function of both the current time $t$ and the time $t_{0}$ when this crosslink was formed during the process.

We shall assume that the
recrosslinking of the dangling chain ends is a simpler case, as the dangling chains are assumed to be in the relaxed state. This is an approximation ignoring the effects of diffusion (possibly reptation) time that is required for this chain to equilibrate in the network. This assumption is also useful in the discussion of the energy of the system, later in the text.
The crosslinking rate can be given by another Kramers expression,
\begin{eqnarray}~\label{rate2}
 \rho_{0}&=& \omega_{0}e^{-w_{c}/k_{B}T},
\end{eqnarray}
where $w_{c}$ is the energy barrier for a dangling chain to overcome in order to be crosslinked.
In this form $\rho_{0}$ is a reaction constant and is independent of the deformation in the system.
Usually, the crosslinking rate is much higher than the breakage rate at ambient temperatures, $\rho_{0}\gg \beta_{0}$ (i.e. $W_b \gg w_c$), so the network can be regarded as `crosslinked'. For high temperatures, one could reach a regime when $\rho_{0}\approx \beta_{0}\approx \omega_{0}$, and this is clearly a system that would undergo a plastic flow under stress. It is interesting that by fitting the data of experiments on vitrimer stretching~\cite{Montarnal,Hillmyer} later in the paper, we shall obtain $W_b \approx 1.4 \cdot 10^{-19}\mathrm{J} = 30\, k_\mathrm{B}T$ at room temperature: a reasonable value much lower than an ordinary covalent bond. 

The rate constant $\rho_0$ measures the reaction time, but we have to also consider the time it would take for the free dangling end of the chain to reach the point of the new crosslinking (a position that we consider force-free for this chain). In some cases, this time is short, e.g. when the crosslinking reaction can happen essentially with any nearby monomer (as happens in vitrimer chemistry~\cite{Montarnal}). In other situations, when the reacting end of a dangling chain needs to travel a substantial distance to link with another matching site, this time can be long.  Many excellent theoretical models describe this diffusion motion (usually -- reptation, with or without constraint release~\cite{Viovy1991,Graham2003}). Here we simply account for the diffusion time as an addition to the reaction time, making the effective rate of re-crosslinking:
\begin{eqnarray}~\label{rate22}
 \rho &=& \frac{1}{t_\mathrm{diff} + 1/\rho_0} , 
\end{eqnarray}
and will later consider the cases when the diffusion time is very short ($t_\mathrm{diff} \gg 1/\rho_0$) and very long ($t_\mathrm{diff} \ll 1/\rho_0$). 

\subsection{Transient network}

Since the crosslinks form and break dynamically,
the numbers of both the crosslinked chains and the dangling chains in the network may change with time.
If we take the number of crosslinked chains at a given time to be $N_{\mathrm{c}}(t)$,
then the number of the uncrosslinked chains is correspondingly $N_{\mathrm{b}}(t)=N_{\mathrm{tot}}-N_{\mathrm{c}}(t)$, where $N_{\mathrm{tot}}$ is the total number of the chains in the system including both crosslinked and freely dangling. If the system is in the equilibrium (reference) state without any deformation,
then the breakage rate in Eq.~\eqref{rate11} becomes a constant $\beta=\beta_{0}e^{\kappa r_{0} } = \beta_{0}e^{3/\sqrt{N_s}} $ (the last relation is due to the average end-to-end in such a network being $\bar{r}_0 = b\sqrt{N_s}$, consistently staying with the Gaussian approximation). 
 The equilibrium detailed balance gives the relationship between $N_{\mathrm{c}}$ and $N_{\mathrm{b}}$ under no deformation: $N_{\mathrm{c}}\beta=N_{\mathrm{b}}\rho_{0}$. Note that it is the reaction rate $\rho_0$, Eq.~\eqref{rate2}, that forms this detailed balance, whereas the full rate, $\rho$, determines the re-crosslinking during the process of dynamic deformation. 

Furthermore,
since the newly re-crosslinked chains are assumed to be in their relaxed state,
the crosslinked chains can be categorized into two classes:
one is the newly crosslinked chains in their force-free relaxed state, with the number $N_{\mathrm{nc}}(t)$, while
the other is the `surviving' crosslinked chains, which were crosslinked initially and are still elastically active at the present time, with the number $N_{\mathrm{sc}}(t)=N_{\mathrm{c}}(t)-N_{\mathrm{nc}}(t)$.

\begin{table}[h]
\newcommand{\tabincell}[2]{\begin{tabular}{@{}#1@{}}#2\end{tabular}}
\begin{tabular}{|c|c|}
\hline
  Time & Number of crosslinked chains \\
\hline\hline
  0 & $N_{\mathrm{c}}(0)$ \\
\hline
$\Delta t$ & $N_{\mathrm{c}}(0)e^{-\beta(\Delta t;0)\Delta t}+N_{\mathrm{b}}(0)\rho \Delta t$\\
\hline
$2\Delta t$ & $N_{\mathrm{c}}(0)e^{-\beta(\Delta t;0)\Delta t}e^{-\beta(2\Delta t;0)\Delta t}+N_{\mathrm{b}}(0)\rho \Delta te^{-\beta(2\Delta t;\Delta t) \Delta t}+N_{\mathrm{b}}(\Delta t)\rho \Delta t$\\
\hline
 ... & ... \\
\hline
$N\Delta t$ & $N_{\mathrm{c}}(0)e^{-\sum_{i=1}^{N}\beta(i\Delta t;0)\Delta t}+\sum_{j=0}^{N-1}N_{\mathrm{b}}(j\Delta t)\rho \Delta t e^{-\sum_{k=j+2}^{N}\beta(k\Delta t;[j+1]\Delta t)\Delta t}$\\
\hline
\end{tabular}
\caption{Time-evolution of the number of the crosslinked chains.}
\label{crossnum}
\end{table}

The Table~\ref{crossnum} illustrates in discrete form how we build up the expressions for the time dependence of $N_\mathrm{c}(t)$. Apart from losing a portion of initially crosslinked chains, at each step with a rate that reflects the current state of deformation, the rate of breaking of newly re-crosslinked chains depends on the changing reference state.  After the first small time interval $\Delta t$,
the number of chains broken from crosslinks is $N_{\mathrm{b}}(\Delta t)=N_{\mathrm{c}}(0)\beta(\Delta t;0)\Delta t$, and the number of the survived crosslinked chains is $N_{\mathrm{sc}}(\Delta t)=N_{\mathrm{c}}(0)(1-\beta(\Delta t;0)\Delta t) \simeq N_{\mathrm{c}}(0)e^{-\beta(\Delta t;0)\Delta t}$, correspondingly.
Meanwhile, the number of the newly crosslinked chains is $N_{\mathrm{nc}}(\Delta t)=N_{\mathrm{b}}(0)\rho \Delta t$. After the next time interval $\Delta t$, the number of the surviving chains initially crosslinked reduces further at a rate  $\beta(2\Delta t;0)$ that corresponds to the state of deformation at this time.
For the chains re-crosslinked at time $\Delta t$, the breakage rate has the reference (force-free) state at $\Delta t$, which explains the second term in the  $2\Delta t$ line of Table~\ref{crossnum}. Plus, a portion of chains that were broken at the previous time step re-crosslinks with the constant rate $\rho$.
Repeating these discrete steps, the total number of crosslinked chains at time $N\Delta t$ can be written down. Taking the limit $\Delta t \rightarrow 0$, the continuous version of  these sums takes the form
\begin{eqnarray}\label{numberof}
 N_{c}(t)=  N_{c}(0) e^{-\int_{0}^{t} \beta(t';0)dt'}+
 \int^{t}_{0}  N_{\mathrm{b}}(t')e^{-\int_{t'}^{t}\beta(t'';t')dt''} \rho \, dt'.
\end{eqnarray}
This expression is key for our subsequent analysis. The first term represents the initially crosslinked chains
surviving from $t'=0$ till the present time,
while the second term represents the chains re-crosslinked during that period both from the originally broken chains and the chains broken at different times during this evolution. Since $ N_{\mathrm{b}}(t) = N_\mathrm{tot} - N_c(t)$, Eq.~\eqref{numberof} is a formal integral equation that determines $N_c(t)$ for a given state of dynamic deformation.

\subsection{Macroscopic elastic energy}

We shall use the classical continuum model of rubber elasticity derived from statistics of Gaussian chains~\cite{Treloar, Doi}. Let us at first assume that a rubbery network, with permanent crosslinks, is at its reference state at $t=0$.
If the system is deformed from its reference state with a general affine deformation tensor $\mathbf{E}(t;0)$ at time $t$,
then the energy density of the system can be written as
\begin{eqnarray}\label{claay}
 F_{\mathrm{rub}}(t;0) &=& \frac{1}{2} G  \left(\mathrm{tr}[\mathbf{E}^{\mathrm{T}}(t;0)\mathbf{E}(t;0)]-3\right) ,
\end{eqnarray}
where $G$ is the shear modulus of the rubber. The entropic Gaussian model will give the rubber modulus proportional to the density of crosslinked chains, $G_0 = k_\mathrm{B}T \, N_c(0)/V$, but we shall not be concerned with a specific value of this material constant.

For a deformed transient network, the average elastic free energy is made of several contributions. Let us assume that the initial reference (force-free) state is at time $t=0$.
For $t>0$, the chains in network no longer have the same reference state: the $N_\mathrm{sc}$ chains crosslinked from the beginning that survived till the current time are deformed with respect to the $t=0$ state, but the $N_\mathrm{nc}$ re-crosslinked chains are deformed with respect to their individual reference states that were force-free at different times.
Consider $N_{\mathrm{nc}}(t_{0})$ chains newly crosslinked at $t_{0}$,
and the macroscopic deformation tensor of the transient network $\mathbf{E}(t_{0};0)$ at time $t_{0}$ (with respect to the original reference state).
{Then $N_{\mathrm{nc}}(t_{0})$ chains are in their reference, or relaxed state, and they do not contribute any elastic energy to the system at time $t_{0}$.
But at a later time, $t>t_{0}$, if the deformation has dynamically changed, the energy density contributed by these $N_{\mathrm{nc}}(t_{0})$ chains is proportional to $N_\mathrm{nc}(t_{0}) e^{-\int^{t}_{t_{0}}  \beta(t';t_{0})dt'} F(t;t_{0})$,
where the time-dependent factor represents the number of the surviving chains which were crosslinked at time $t_{0}$.} The elastic free energy density $F(t;t_0)$ in this expression is determined by the deformation tensor $\mathbf{E}(t;t_{0})$ with respect to the reference state at $t_0$, expressed by
\begin{eqnarray}\label{tenso}
 \mathbf{E}(t;t_{0})&=& \mathbf{E}(t;0) \cdot \mathbf{E^{-1}}(t_{0};0),
\end{eqnarray}
where $\mathbf{E^{-1}}$ is the inverse matrix of $\mathbf{E}$.

Assembling together all these contributions from the chains that have been re-crosslinked during the deformation period between $t'=0$ and $t$, and adding the continuously diminishing contribution from the initially crosslinked chains, the
 energy density of the transient network can be expressed by
\begin{eqnarray}\label{energy}
F_{\mathrm{tr.n.}}(t) &=& e^{-\int^{t}_{0}\beta(t';0)dt'}F_{\mathrm{rub}}(t;0)
         +\int^{t}_{0}\rho \, \frac{N_{\mathrm{b}}(t')}{N_{c}(0)}e^{-\int_{t'}^{t}\beta(t'',t')dt''}F_{\mathrm{rub}}(t;t')dt' ,
\end{eqnarray}
where in the second term the neo-Hookean free energy density uses the dynamically changing strain tensor from Eq.~\eqref{tenso}.

In ordinary rubbery networks, the crosslinks are permanent, and the rubber modulus $G$  is defined in Eq.~\eqref{claay} with an unchanged reference state at $t=0$.
However, the reference state in a transient network can only be defined locally for different chains,
depending on when they are crosslinked. Because of the difficulty in tracking the real reference state of every crosslinked chain, it is sometimes convenient to define an effective shear modulus $G^{*}$ as the ratio~\cite{Mayumi2013}
\begin{eqnarray}\label{modul}
 G^{*}(t)=\frac{2F_{\mathrm{tr.n.}}(t)}{\mathrm{tr}[\mathbf{E}^{\mathrm{T}}(t;0)\mathbf{E}(t;0)]-3} ,
\end{eqnarray}
which essentially measures the relative change of the transient network response with respect to an analogous permanently crosslinked network with the elastic reference state at $t=0$.

\subsection{Elastic stress tensor}

The stress of a transient network usually includes two parts, the elastic stress and the viscous stress, $\mathbf{\sigma}^{\mathrm{ela}}+\mathbf{\sigma}^{\mathrm{vis}}$, since the plastic flow could be an essential part of the mechanical response. The origins of the viscous part $\mathbf{\sigma}^{\mathrm{vis}}$ are complex, and might include nonaffine movement of the crosslinks, dynamics of entanglements and dangling chains, etc. We shall simply express it in the form
\begin{eqnarray}
  \mathbf{\sigma}^{\mathrm{vis}} &=& \mathbf{\eta}(\mathbf{\dot{\gamma}})\cdot \mathbf{\dot{\gamma}} ,
\end{eqnarray}
where $\mathbf{\eta}$ is the viscosity tensor, which is expressed as a possible function of the strain rate tensor $\mathbf{\dot{\gamma}}$.
There are many studies on how viscous stress depends on the strain rates,
including shear thinning and thickening effects, which is usually induced by nonaffine movement inside of the network~\cite{Indei2005, Koga2005}.

In this work we will concentrate on how elastic stress evolves with deformations ignoring the viscous effects during the developed plastic flow.
Earlier we discussed the Helmholtz elastic free energy of the transient network. However, we need to account for the material (in)compressibility, which is not naturally included in the classical rubber-elasticity expression \eqref{claay}. It is common to simply impose the incompressibility constraint onto such an expression; however, the `cost' is often an unphysical non-zero stress on the free sides of the deformed sample. There are two ways to account for this: either explicitly include the (large) bulk modulus, find a corresponding (small) volume change on deformation and rescale the strain tensor to be measured with respect to that state~\cite{Wendy2005} -- or work with the Gibbs free energy density $g(p,T)$ and replace the (constant) pressure from the constraint that free surfaces of the sample have zero stress~\cite{Doi2005}. This is the approach we follow here and introduce:
\begin{eqnarray}~\label{g}
g(t)=F_{\mathrm{tr.n.}}(t)-p \cdot\mathrm{det}\mathbf{E} \ ,
\end{eqnarray}
where $F_{\mathrm{tr.n.}}(t)$ is given by Eq.~\eqref{energy}, $E_{\mathrm{ij}}(t;t')=E_{\mathrm{ik}}(t;0)E_{\mathrm{kj}}^{-1}(t';0)$, and the pressure $p$  is a Lagrangian multiplier in charge of the incompressibility condition, determined by the boundary conditions of the stress.
Defining stress as a functional variation of $g(t)$,
\begin{eqnarray}
 \sigma^{\mathrm{ela}}_{\mathrm{ij}}(t) &=& \frac{\delta g(t)}{\delta E_{\mathrm{ij}}(t;0)},
\end{eqnarray}
we can obtain the expression of the stress tensor,
\begin{eqnarray}
 \sigma_{\mathrm{ij}}^{\mathrm{ela}}(t) &=& e^{-\int^{t}_{0}\beta(t';0)dt'}G \, E_{\mathrm{ij}}(t;0)
         +
\int^{t}_{0}\rho  \, \frac{N_{\mathrm{b}}(t')}{N_{c}(0)}e^{-\int_{t'}^{t}\beta(t'',t')dt''}G \, E_{\mathrm{ik}}(t;t')E_{\mathrm{jk}}^{-1}(t';0)dt' \nonumber \\
&&-p\cdot\mathrm{det}\mathbf{E}\cdot E_{\mathrm{ji}}^{-1},  \label{stressyy}
\end{eqnarray}
where the first term represents the contribution from the surviving chains crosslinked at $t=0$,
and the second term represents the contribution from the chains re-crosslinked between $t'=0$ and $t$.

\begin{figure} [t]
\begin{center}
\includegraphics[width=0.6\columnwidth]{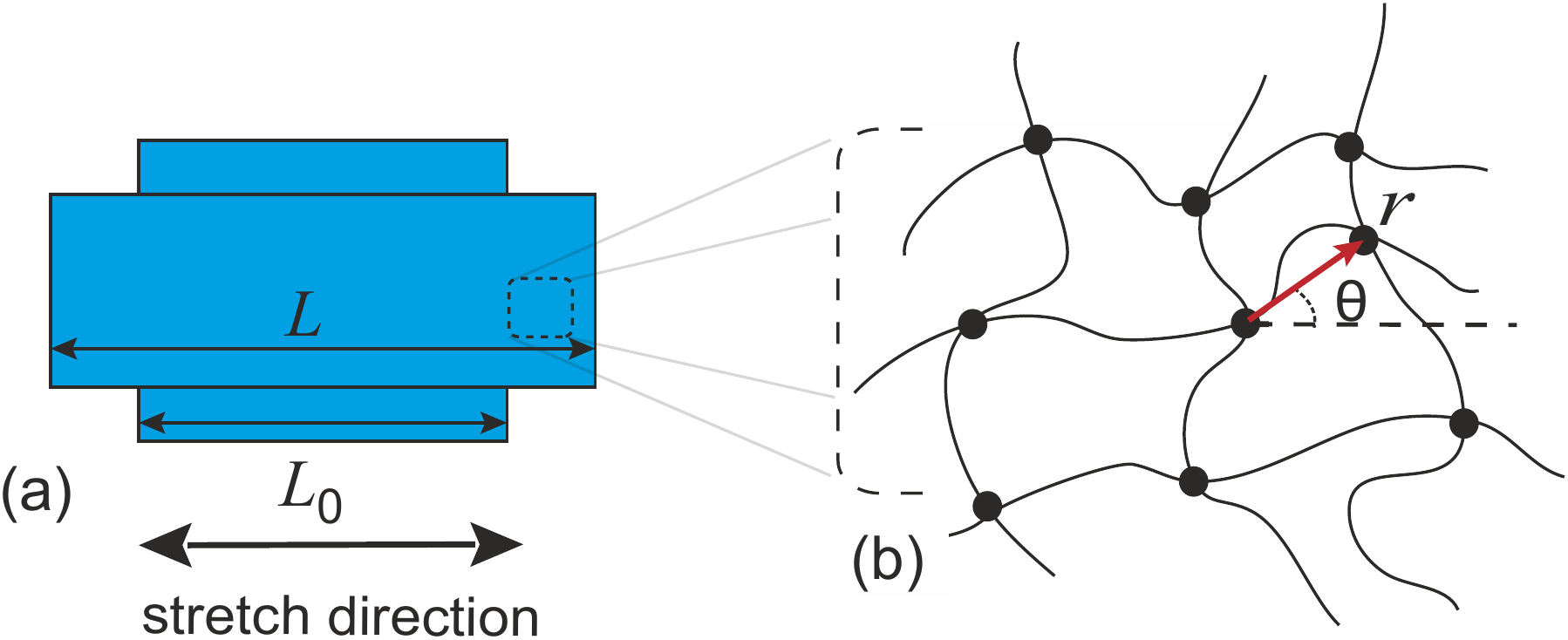}
\caption{Schematic illustration of (a) a polymeric sheet, and (b) a chosen subchain in a uniaxial stretched network,
with $r$ as the end-to-end distance and $\theta$ as the angle between the end-to-end vector and the stretch direction.}
\label{stretch}
\end{center}
\end{figure}

Let us now focus on how a transient  network responds to an imposed uniaxial stretch, as an application of the above general model. When undergoing a uniaxial stretch along the longitudinal direction, Fig.~\ref{stretch}(a),
the polymeric sheet will deform,
with length as $L=\lambda_{\mathrm{L}} L_{0}$, width as $W=\lambda_{\mathrm{W}}W_{0}$ and thickness as $H=\lambda_{\mathrm{H}}H_{0}$,
where $\lambda_{\mathrm{L}}$, $\lambda_{\mathrm{W}}$, $\lambda_{\mathrm{H}}$ are elongation ratios along the three orthogonal directions. Taking $\lambda_{\mathrm{L}}$ as the external parameter $\lambda$, $\lambda_{\mathrm{W}}$ and $\lambda_{\mathrm{H}}$ can be written as $1/\sqrt{\lambda}$ each, due to the incompressibility.

If the particular crosslinks are formed at time $t'$, then their corresponding deformation tensor at time $t$ can be known from Eq.~\eqref{tenso}, treating $\mathbf{E}(t';0)$ as the reference state:
\begin{eqnarray}  \label{uniE}
  \mathbf{E}(t;t')  &=&\frac{\lambda(t)}{\lambda(t')}\mathbf{e}_{\mathrm{L}}\mathbf{e}_{\mathrm{L}}+
  \sqrt{\frac{\lambda(t')}{\lambda(t)}}\left(\mathbf{e}_{\mathrm{W}}\mathbf{e}_{\mathrm{W}}+
  \mathbf{e}_{\mathrm{T}}\mathbf{e}_{\mathrm{T}}\right),
\end{eqnarray}
where $\mathbf{e}_{\mathrm{L}}$, $\mathbf{e}_{\mathrm{W}}$ and $\mathbf{e}_{\mathrm{T}}$ are unit vectors along the three orthogonal directions. In this case, Fig.~\ref{stretch}(b), the average end-to-end distance $\langle r \rangle$ that determines the breaking rate $\beta$ in Eq.~\eqref{rate11} can be calculated using the changing average end-to-end distance that reflects the deformation that occurs at time $t$ with respect to a reference state at time $\tau$:
\begin{eqnarray}~\label{rate1_2}
 \langle r_{t;\tau}\rangle&=&
r_{0} \int_{0}^{\pi/2}d\theta  \sin\theta\sqrt{\left(\frac{\lambda(t)}{\lambda(\tau)}\right)^{2}\cos^{2}\theta + \frac{\lambda(\tau)}{\lambda(t)}\sin^{2}\theta}.
\end{eqnarray}
where $r_{0}\sim \sqrt{N_{s}}b$ is the mesh size of the network in its reference state.
Substituting the strain tensor from Eq.~\eqref{uniE}, the Helmholtz elastic free energy density of the system can be written explicitly as:
\begin{eqnarray}\label{energyyy}
F_{\mathrm{tr.n.}}(t) &=& \frac{1}{2} G\, e^{-\int^{t}_{0}\beta(t';0)dt'}\left(\lambda(t)^{2}+\frac{2}{\lambda(t)}-3\right)   \\
         &&+\frac{1}{2}G \int^{t}_{0}
         \rho \, \frac{N_{\mathrm{b}}(t')}{N_{0}}e^{-\int_{t'}^{t}\beta(t'',t')dt''}
         \left[\left(\frac{\lambda(t)}{\lambda(t')}\right)^{2}+
         \frac{2\lambda(t')}{\lambda(t)}-3\right]dt'    \nonumber
\end{eqnarray}
with the orientational averaging implicit in the expressions for $\beta(t,t')$ in the relaxation exponents. 
Applying Eq.~\eqref{modul}, the effective shear modulus can be obtained by simply dividing both terms in this free energy density by the characteristic neo-Hookean strain combination, which for uniaxial deformation is given by the bracket in the first term in  Eq.~\eqref{energyyy}:
\begin{eqnarray}~\label{o_shear_modulus}
G^{*}(t)&=&G\, e^{-\int^{t}_{0}\beta(t';0)dt'}
     \\&&+
    G \int^{t}_{0} \rho \, \frac{N_{\mathrm{b}}(t')}{N_{0}}e^{-\int_{t'}^{t}\beta(t'',t')dt''}
         \left(\frac{\lambda(t)^{2}/\lambda(t')^{2}+
         2\lambda(t')/\lambda(t)-3}{\lambda(t)^{2}+2/\lambda(t)-3}\right)dt' .     \nonumber
\end{eqnarray}
The transverse diagonal components of stress  can be obtained from Eq.~\eqref{stressyy} by inserting the explicit components of the uniaxial strain tensor, producing
\begin{eqnarray}
  \sigma_{\mathrm{W}}=\sigma_{\mathrm{T}}&=& \frac{G}{\sqrt{\lambda(t)}} \left( e^{-\int^{t}_{0}\beta(t';0)dt'}
         +\int^{t}_{0}\frac{N_{\mathrm{b}}(t')}{N_{0}}\rho \, e^{-\int_{t'}^{t}\beta(t'',t')dt''}\lambda(t')dt'\right)-p \sqrt{\lambda(t)}.
\end{eqnarray}
In this geometry of uniaxial stretching, $\sigma_{\mathrm{W}}$ and $\sigma_{\mathrm{T}}$ should be both equal to  $0$,
which gives the value of $p$ to be substituted into the final expression for the tensile stress. After a little algebra we obtain:
\begin{eqnarray}\label{stress_uni}
 \sigma_{\mathrm{L}} (\lambda, t) &=& G \, e^{-\int^{t}_{0}\beta(t';0)dt'} \left(\lambda(t)-\frac{1}{\lambda(t)^{2}}\right)
\\ && +G \int^{t}_{0} \frac{N_{\mathrm{b}}(t')}{N_{0}}\rho \, e^{-\int_{t'}^{t}\beta(t'',t')dt''}\left(\frac{\lambda(t)}{\lambda(t')^{2}}-\frac{\lambda(t')}{\lambda(t)^{2}}\right)dt'. \nonumber
\end{eqnarray}
Calculation of this dynamic stress for a given imposed deformation $\lambda(t)$ goes in two steps: first we must solve the integral equation \eqref{numberof} to determine $N_\mathrm{b}(t)$ and then compute the time-integrals in Eq.~\eqref{stress_uni}.  In the following sections
we will discuss in detail how a transient network responds to
several practically relevant deformation modes: step strain, ramp deformation, and a loading-unloading cycle.

\section{Stress relaxation}
In this section, we discuss how the stress in a transient network relaxes in a `standard experiment' when a uniaxial stepwise deformation $\lambda_{\mathrm{L}}=\lambda$ is applied at $t=0$. This is the simplest case of application of our theory. 
As seen in Eq.~\eqref{o_shear_modulus}, with $\lambda(t)=\lambda(t')$ the second term vanishes exactly, which means the chains re-crosslinked after $t=0$ do not contribute to the relaxation stress,
as these chains remain in their force-free reference state with $\lambda(t)=\lambda$.
From Eq.~\eqref{stress_uni}, we can directly find the tensile stress along the stretching direction, which relaxes as a simple exponential:
\begin{eqnarray}\label{stepstrain}
 \sigma_{\mathrm{L}} &=&G \, e^{-\beta(\lambda) t} \left(\lambda-\frac{1}{\lambda^{2}}\right),
\end{eqnarray}
where the inverse $\tau=1/\beta (\lambda)$ is the characteristic relaxation time of the tensile stress~\cite{Long2013, Yu2014}.
The explicit form of $\beta(\lambda) = \beta_0 \exp [\kappa \langle r(\lambda) \rangle ]$  with the orientational average of the end-to-end chain length from Eq.~\eqref{rate1_2} is given by:
\begin{eqnarray}~\label{rate215}
 \beta(\lambda)&=&\omega_{0}e^{\kappa r_{0}\int_{0}^{\pi/2}
 \sin\theta\sqrt{\frac{1}{\lambda}\sin^{2}\theta+\lambda^{2}\cos^{2}\theta}d\theta}e^{-W_{\mathrm{b}}/k_{\mathrm{B}}T}
 =c_{0}(\lambda)e^{-W_{\mathrm{b}}/k_{\mathrm{B}}T}, 
\end{eqnarray}
where
\begin{eqnarray}
 c_0(\lambda) = \omega_0 \exp \left[ \frac{3}{2 \sqrt{N_s \lambda}} \left( \lambda^{3/2} + \frac{ \mathrm{Arcsinh} \sqrt{\lambda^3-1}}{\sqrt{\lambda^3-1}}  \right)   \right] , \nonumber 
\end{eqnarray}
which increases monotonically with the stretching ratio $\lambda$ (and also on uniaxial compression, $\lambda < 1$). At small strain $\varepsilon = \lambda-1 \ll 1$, we obtain $c_0 \approx \omega_0 \exp (3/\sqrt{N_s})$, a constant for a given network. For large $\lambda$, the opposite limiting case gives $c_0 \approx \omega_0 \exp (3 \lambda /2\sqrt{N_s})$, that is, the rate of breaking increases exponentially. In this case most of the chains align along the stretching direction and directly transmit the deformation to the shift in the thermal activation law. 

\begin{figure}[t]
\centering
\includegraphics[width=0.95\columnwidth]{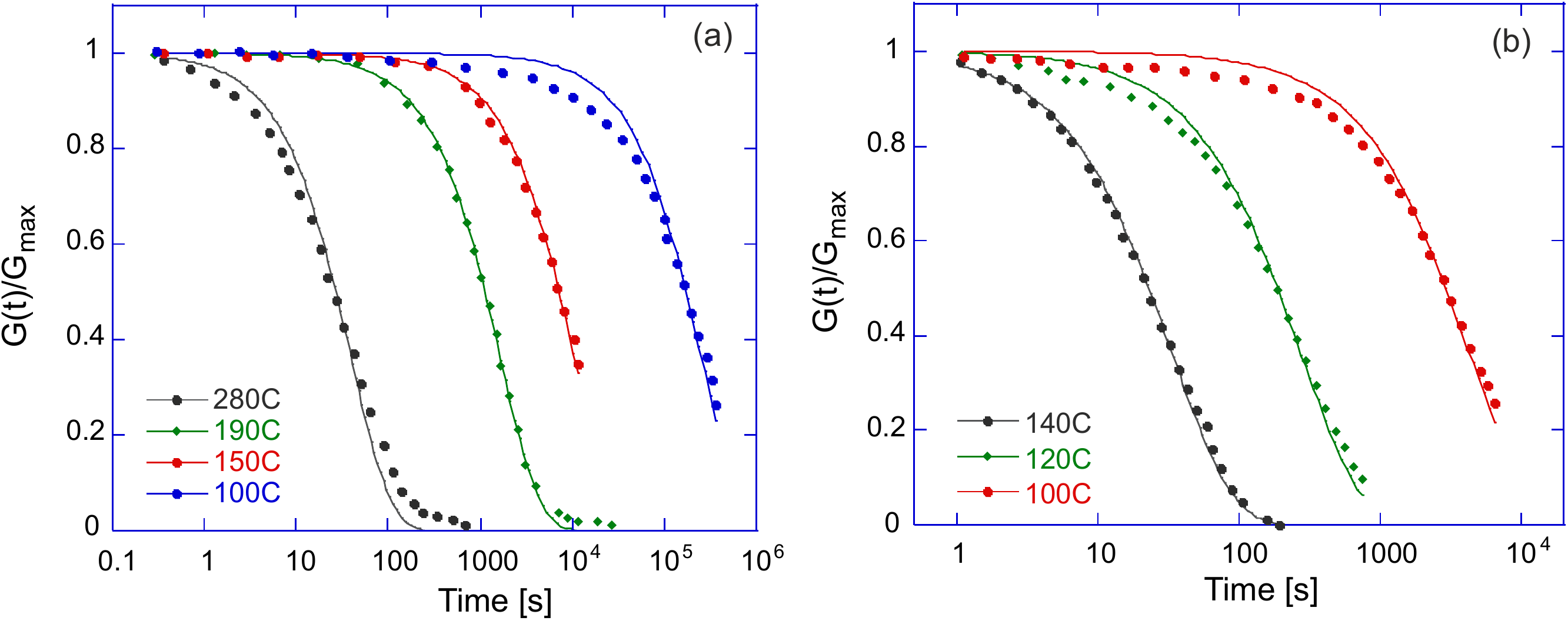}
\caption{(a) Relaxation of the effective shear modulus $G^{*}$ for different temperatures in two vitrimer networks.
Solid lines are the simple exponential curves, and the dots are experimental data:  (a) from Leibler \textit{et al.}~\cite{Montarnal}, where the fitting gives  $W_\mathrm{b} \approx 1.4\cdot 10^{-19} \, \mathrm{J} =34~k_\mathrm{B}T_\mathrm{room}$.  (b)  A different polylactide vitrimer from Hillmyer \textit{et al.}~\cite{Hillmyer} gives a much stronger bonding: $W_\mathrm{b}  \approx 2.6\cdot 10^{-19} \, \mathrm{J} =64~k_\mathrm{B}T_\mathrm{room}$. }
\label{Grelax}
\end{figure}

Most standard stress-relaxation experiments are conducted in the linear stress-strain regime, effectively measuring the effective shear modulus $G^{*}(t)$. Figure~\ref{Grelax} shows two examples of analysis of experimental data in two chemically different vitrimer networks, assuming that in both cases the authors did maintain the linear stress-strain regime. Both plots show that the simple exponential relaxation is a valid model, and since the data at different temperatures has been collected -- we can fit the Arrhenius law in Eq.~\eqref{rate215} and obtain the activation energies $W_\mathrm{b}$ for the transesterification reaction in these two materials (the values listed in the figure caption).

\begin{figure} [t]
\centering
\includegraphics[width=0.95\columnwidth]{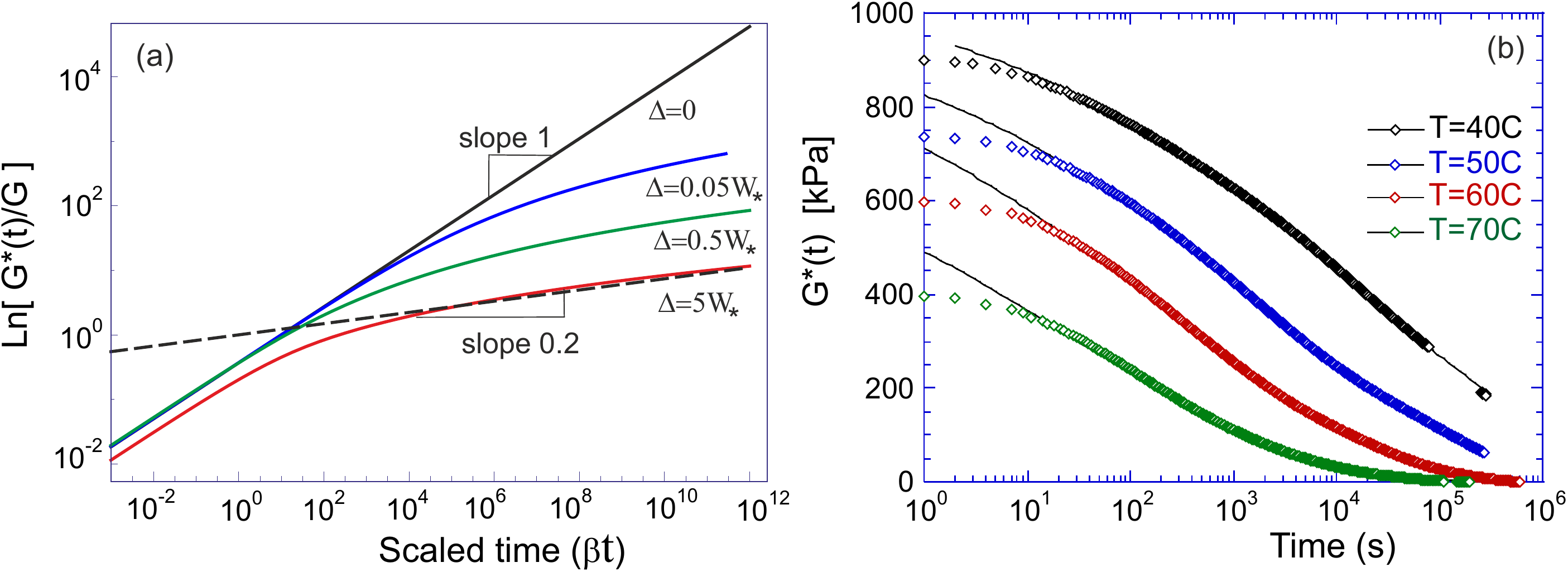}
\caption{(a) Double log-log plots of the Eq.~\eqref{quench}, in scaled non-dimensional variables, for several values of variance (width) $\Delta$ of the quenched distribution of energy barriers $W_\mathrm{b}$. The dashed line has a slope of 0.2, giving the long-time relaxation limit of $\langle G^* \rangle_W \propto \exp[-(\beta t)^{0.2}]$ after the crossover from the linear-exponential regime at early times. 
(b) Relaxation of the effective shear modulus $G^*(t)$ for different temperatures in the transient network of SIS. Here the solid lines are the stretched exponential curves $\exp[-(\beta t)^{0.2}]$ resulting from our model with a broad  distribution of activation energies $W_\mathrm{b}$, and the dots are experimental data from Hotta \textit{et al.}~\cite{Hotta2002} Clearly the stretched exponential fits the long-time relaxation while the short-time process is different.}
\label{Grelax2}
\end{figure}

So far we worked under assumption that the activation energy for the crosslink breaking, $W_\mathrm{b}$, is a fixed parameter of the material. This is a good assumption in the case when the crosslinks are held by, e.g. hydrogen bonds, or in the case of vitrimers (where the covalent bond is `weakened' by an appropriate catalyst). However, there are many cases where the physical bonds would not have a single characteristic binding energy: the simple example is the SIS telechelic block-copolymer network where the glassy polystyrene micelles must have a distribution of sizes, shapes, and therefore strength of chain confinement. The way to account for such a distribution is to perform the quenched average of the relaxation function \eqref{stepstrain} with an (assumed Gaussian) probability distribution:
\begin{eqnarray}\label{quench}
 \langle G^*(t) \rangle_W = G \, \int \exp \left[ -\omega_0 e^{\kappa r_0} e^{-W_\mathrm{b} / k_\mathrm{B}T} \, t \right]  \cdot \sqrt{\frac{\Delta}{2\pi}} \, e^{-(W_\mathrm{b} - W_*)^2/2\Delta} \ dW_\mathrm{b} ,
\end{eqnarray}
where $W_*$ is the average binding energy and $\Delta$ measures the spread of the distribution. The earlier case of the single binding energy is $\Delta \rightarrow 0$. The integral of the double exponential is difficult to calculate analytically (although good interpolations are possible), but the numerical plot of the quenched-averaged relaxation function $ \langle G^*(t) \rangle_W$ in Fig.~\ref{Grelax2}(a) shows that the relaxation law becomes the stretched exponential $\exp [-(\beta t)^{0.2}] $ when there is a sufficiently wide spread of the $W_\mathrm{b}$ values: $\Delta \geq W_*$, while remaining the simple exponential for the narrow distribution, as expected.  Also note that this characteristic stretched exponential only sets in at long relaxation times, while the short-time remains simple exponential, with the crossover between the two regimes starts at times  $(\omega_0 e^{\kappa r_0}) t \sim 1$. 
The relaxation data in Fig.~\ref{Grelax2}(b) are from the physically crosslinked SIS elastomer of Hotta \textit{et al.}~\cite{Hotta2002} where the long-time tails are reliably following the $\exp [-(\beta t)^{0.2}] $ law, supporting the concept of a broad distribution of crosslinking strengths in such a physically linked network.

\begin{figure} 
\centering
\includegraphics[width=0.45\columnwidth]{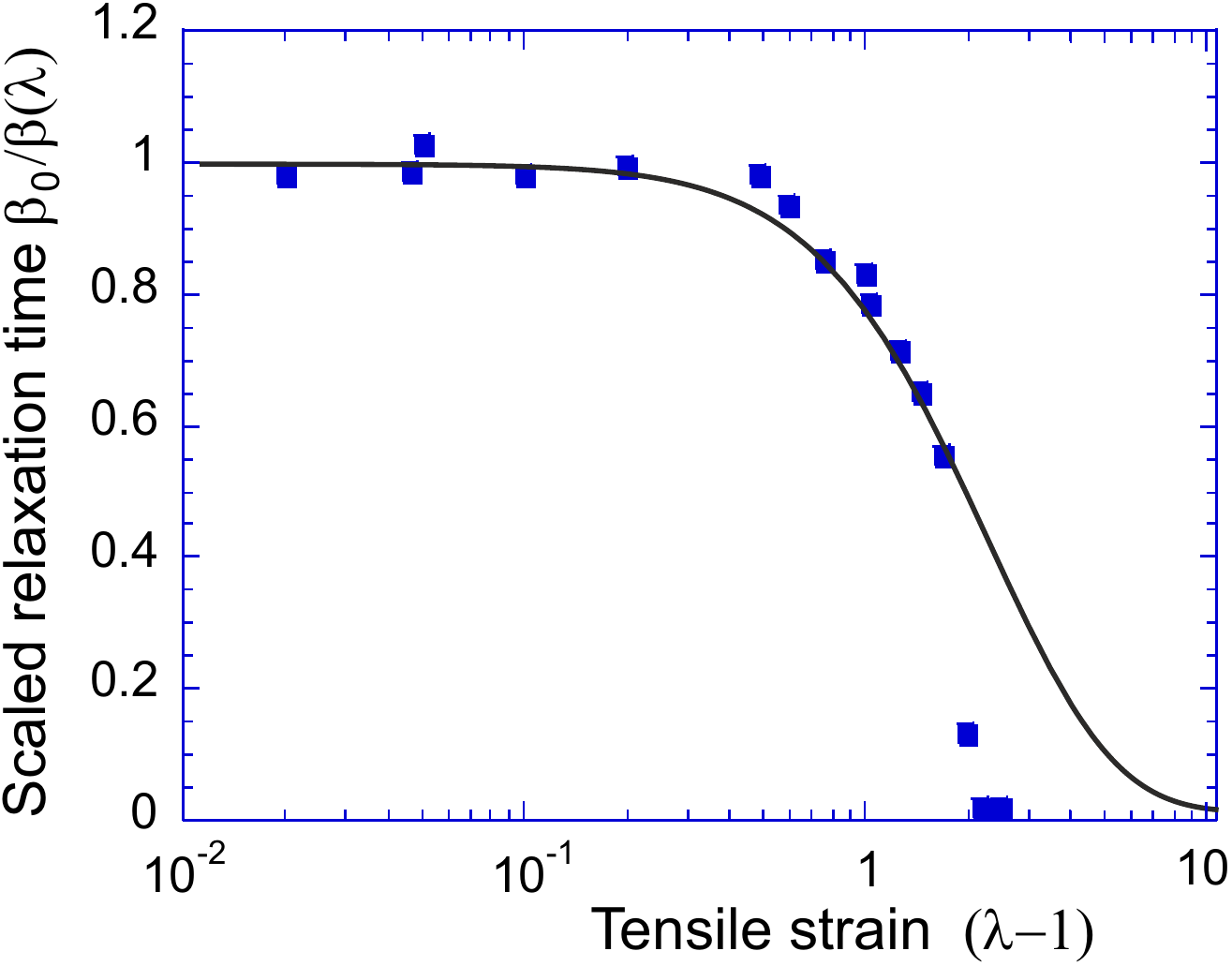}
\caption{ Relaxation time $1/\beta(\lambda)$ in the non-linear regime, plotted as a function of the strain $\varepsilon=\lambda-1$.
Solid line is the theoretical result of Eq.~\eqref{rate215}, and the dots are experimental data from Serero \textit{et al.}~\cite{Serero2000} The single fitted parameter here is $\kappa r_{0} \approx 1.7$.  The deviation from the theory at high strain is certainly due to the sample tearing. }
\label{Grelax3}
\end{figure}

There are very few papers where the stress relaxation in transient networks is experimentally studied at increasing magnitude of the step strain $\lambda$, with the work of Serero \textit{et al.}~\cite{Serero2000} being one of the few. Although the stretched exponential $G^{*}=G\, e^{-(\beta t)^{0.8}}$ was used in, the results would be qualitatively the same with what we get in above case of a simple exponential. We find that the experimental values for $\beta(\lambda)$ fit very well with the full high-strain expression in Eq.~\eqref{rate215}. 

\section{Strain ramp}

\begin{figure} [t]
\centering
\includegraphics[width=0.95\columnwidth]{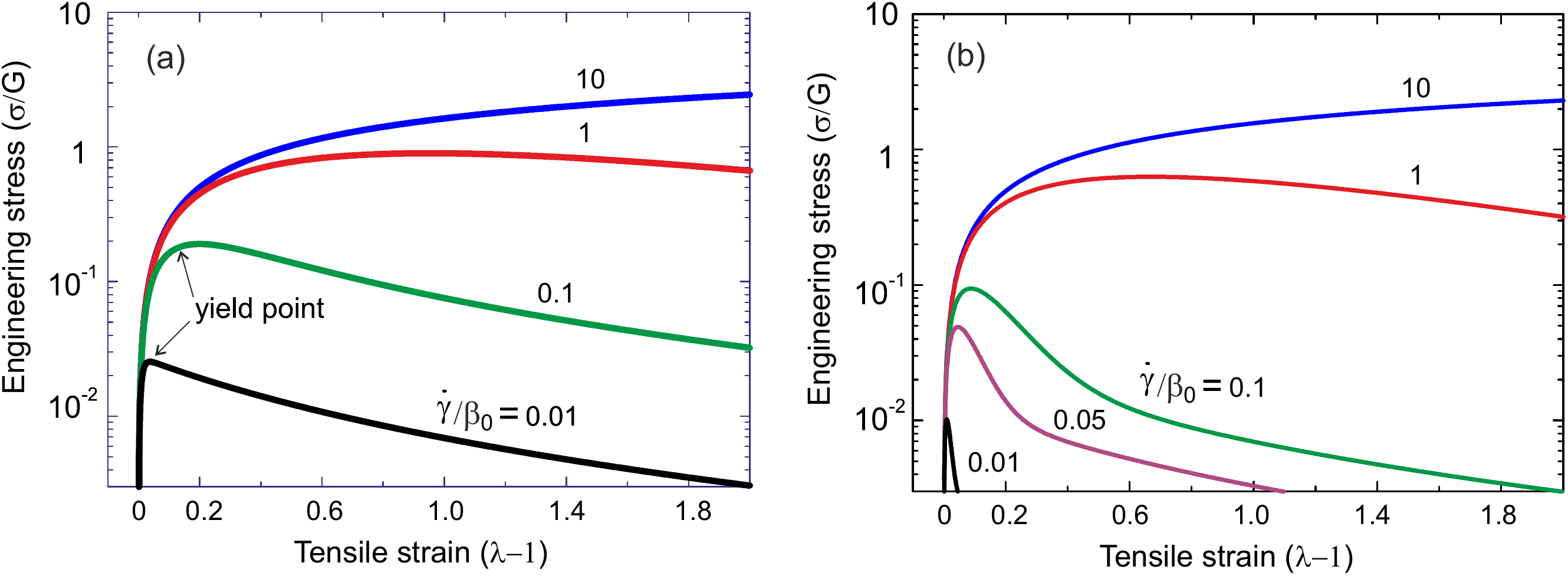}
\caption{Strain-stress relations of a transient network under a linear ramp deformation for different strain rates, with plot (a) showing the case of fast re-crosslinking: $\rho = 10 \beta_0$, and plot (b) the slow re-crosslinking, $\rho = 0.1 \beta_0$.}
\label{ram}
\end{figure}

The other commonly used testing method in rheology is the linear ramp of imposed strain. Many standard instruments, such as Instron, operate in this mode, and very often one finds the stress-strain curves in the literature are reported after measuring the strain as a function of time during a strain ramp. Here we analyze how the dynamics of crosslink distribution shows itself in such an experiment. We remain in the uniaxial stretching geometry and let the longitudinal extensional strain increase linearly with time, $\lambda=1+\dot{\gamma} t$,
where $\dot{\gamma}$ is a constant strain rate.
We already know the dynamic strain-stress relationship in the uniaxial geometry, which is Eq.~\eqref{stress_uni}, so all we need is to identify the important non-dimensional parameters that control the outcome.  Let us measure the time in units of $1/\beta_0$, and similarly for the strain rate, $\dot{\gamma}/\beta_0$, and consider two cases: of fast re-crosslinking, $\rho = 10 \beta_0$, and slow re-crosslinking, $\rho = 0.1 \beta_0$ (meaning that the diffusion time $t_\mathrm{diff}$ is long in the second case). Then, measuring the stress in units of raw rubber modulus $G$, we can numerically integrate Eq.~\eqref{stress_uni} and plot the results in Fig.~\ref{ram}. 

We see that initially the stress increases linearly with elongation ratio $\lambda$ (or strain $\lambda-1$),
and the slope is exactly the shear modulus $G$.
There is always a point of `stress overshoot' (the yield point~\cite{Groot1996}) for every $\dot{\gamma}$, although at very fast rates of deformation this point moves far to the right in the plots. Past this yield point the stress begins to monotonically decrease with strain, with a power-law numerically found close to $\lambda^{-2}$. 

The phenomenon of `stress overshoot' is encountered often in rheological studies of disordered materials, and the detailed mechanisms vary for different systems.
In entangled polymer solutions and polymer melts, the
Doi-Edwards-Marrucci-Grizzutti model predicts the existence of stress overshoot~\cite{Pearson, Mead},
which originates from the contraction of stretched chains and reptation of polymer chains in the tubes.
Later, the idea of "constraint release" was proposed and developed~\cite{Viovy1991, Colby1990, Milner2001, Pattamaprom2001, Graham2003} to produce an even more pronounced stress overshoot and yielding instability. One also finds stress overshoot in metallic glass~\cite{Zaccone2014, Kato1998, Lu2003}, where the softening and fluidization is prompted by the nonaffine shear-induced cage breakup. One finds a lot of conceptual similarity in all these physical situations, where the conditions are reached to break the microscopic constraints that normally produce an elastic contribution. 

\begin{figure} [t]
\centering
\includegraphics[width=0.95\columnwidth]{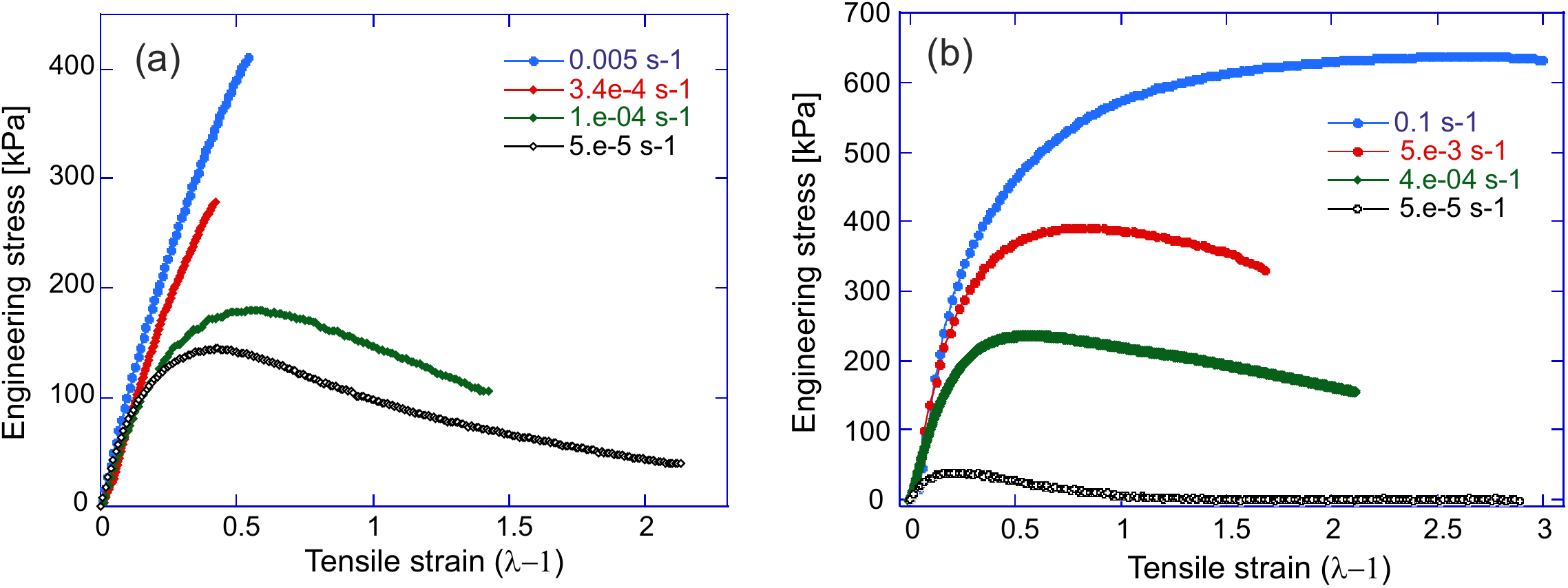}
\caption{Strain-stress relations of a transient network under a linear ramp deformation for different strain rates, with plot (a) showing the data for the vitrimer of Leibler \textit{et al.}~\cite{Montarnal}, at constant temperature $T=130^\circ$C, and plot (b) the data for the physically crosslinked SIS network of Hotta \textit{et al.}~\cite{Hotta2002}, at constant temperature $T=80^\circ$C. In both cases the temperature is chosen at the approximate level of `vitrification transition'; the rates of strain are labelled on the plots. }
\label{ram2}
\end{figure}

To test the predictions of our theory, we carried out strain-ramp experiments on two very different transient networks: the classical vitrimer and the physically crosslinked SIS elastomer, Fig.~\ref{ram2}. We used the custom-built mechanical testing gear described elsewhere~\cite{Robyn}, which in this situation has been set to impose a constant controlled rate of uniaxial extension on the sample, while continuously monitoring its tensile stress and changes in shape. In order to find the stress overshoot within the comfortable range of strain rates and stress values, we had to maintain the temperature close to the vitrification point, as defined for both materials in the original paper~\cite{Montarnal,Hotta2002}, respectively. In full agreement with theoretical curves in Fig.~\ref{ram}, the experiment on both materials shows a clear yielding instability and the continuous decrease of stress past it, when the rate of stretching is sufficiently low. The vitrimer network was not able to survive without fracturing at higher strain rates, while the SIS (with its generally more robust composite microstructure and longer chain strands) shows the high-rate curves also in agreement with Fig.~\ref{ram}.

\begin{figure} [t]
\centering
\includegraphics[width=0.5\columnwidth]{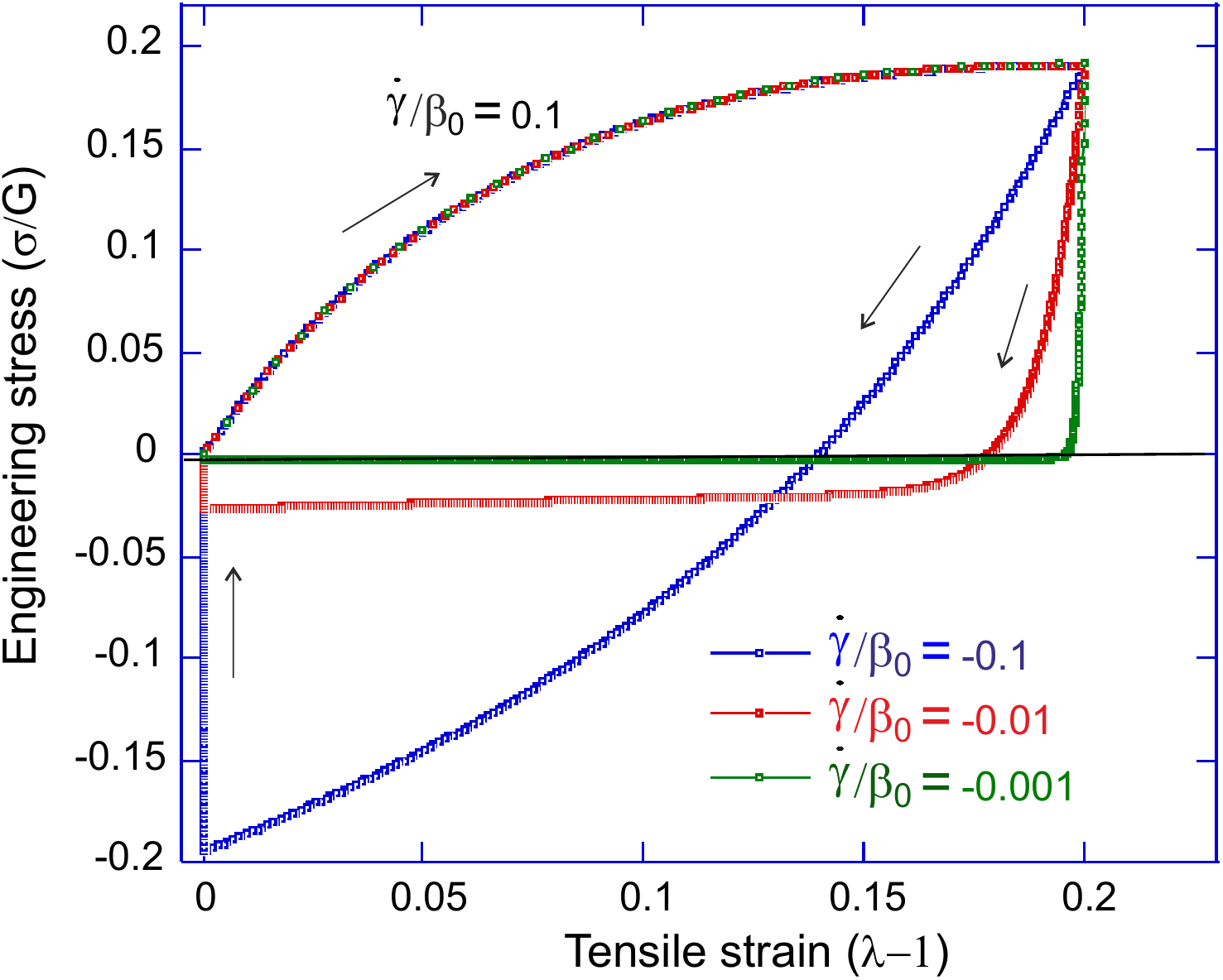}
\caption{Strain-stress relation for several loading-unloading cycles, in all cases with loading rate $\dot{\gamma}/\beta_0 = 0.1$ and several unloading (negative ramp) rates labelled on the plot;  $\rho = 10 \beta_0$. Once the (imposed) sample length returns to its original value, it is held fixed for the period of stress relaxation. The fact that the next loading cycle follows exactly the same curve indicates the full recovery of the sample reference state.}
\label{loadunloa}
\end{figure}

Self-healing materials attract much attention due to their potential applications in mimicking biological tissues,
advanced materials with reversible performance, and in the general context of re-using recycled plastic components. One of the aspects of self-healing is the reproducibility of repeated stretching cycles. Both the stretching and the return to the original imposed length are assumed to proceed as a linear ramp with the strain rate $\dot{\gamma}$. The dynamic tensile stress response is still given by Eq.~\eqref{stress_uni}, and Fig.~\ref{loadunloa} illustrates the response over a sequence of deformation cycles, taking a constant rate of loading that corresponds to the `0.1' curve in Fig.~\ref{ram}(a) reaching just before the yield instability point, followed by a constant rate of unloading-compression. Several rates of unloading are presented to illustrate the dynamics of the process, but in each case the tensile stress passes the zero point and turns into compression when the length of the sample is forced to shorten. The negative (compression) stress reaches the maximum magnitude when the stress returns to zero, at which point we hold the shape constant for a period of relaxation. In fact, this stress relaxation under an effective compression step is not different from the one studied in Fig.~\ref{Grelax} and Eq.~\eqref{stepstrain}: it is a simple exponential relaxation over a characteristic time $\beta_0 t \approx 1$ for all three unloading curves -- only the amplitude of stress changes at different rates. 

In Fig.~\ref{loadunloa}, we see that the compression stress is larger for the same stretching ratio if the unloading rate is higher, which is because fewer stretched chains are able to relax or disconnect from the stretched crosslinks. Obviously, more elastic energy is relaxed or dissipated with a lower unloading rate,
due to relatively quick breakage and reformation of new crosslinks.
However, in such a loading-unloading experiment, a significant practical factor might be the Euler buckling of the elastomer sample on compression~\cite{Landau, Feynman}. The bucking instability occurs when a compression force on a rod of length $L$ exceeds the critical value $f_c=\pi^2 B/L^2$, where $B$ is the bending modulus. Assuming the rectangular cross-section of the sample with the width $W$ and thickness $H$, this modulus is $B=3G\, WH^3/12$ and the critical stress is $\sigma_c = f_c /WH$. We then find the critical compression stress at which the sample would buckle: $\sigma_c = \frac{1}{4}\pi^2 G (H/L)^2$. So for a typical sample in a shape of flat strip, with $H/L \ll 1$, the negative (compression) values of stress in Fig.~\ref{loadunloa} are not achievable. Instead, the sample would buckle very soon on entering the compression region, and the `recovery' we observed in these plots will not be possible. Nevertheless, the concept of self-healing remains valid: on applying a required set of constraints (in shape or stress) the transient network can be brought into any desired reference state.

\section{Conclusion}

In this work, we have derived the dynamic constitutive relation of a transient network, in which crosslinks can be broken by local tensile force on the polymer strand connecting them -- and re-established in the assumed zero-stress configuration with a certain rate. To achieve this, we had to combine the microscopic kinetic description of crosslinks with the macroscopic rubber-elastic energy function describing the deviation from the dynamically changing reference state. The incompressibility constraint is accounted for via the pressure acting as Lagrange multiplier, ensuring the boundary condition constraints are satisfied. 

After the general analysis, we specifically focus on the case of uniaxial deformation and the main Eq.~\eqref{stress_uni} is the constitutive relation for that case. There are two particular applications we consider: the relaxation of stress after a static imposed strain, and the response to a dynamic strain imposed as a constant-rate ramp (in the latter case, also the cyclic loading-unloading deformation). In both cases we compare the detailed theoretical predictions with experimental results: obtained from the literature in the case of stress relaxation, and our own in the case of dynamic loading. In both cases we compare two very different kinds of transient network: the SIS tri-block copolymer physically bonded via phase-separated glassy micelles, and the vitrimer networks where the covalent bonds can be reconfigured by the transesterification reaction. 

The most important conclusion about the stress relaxation is that it proceeds in an exponential manner. This is in marked contrast to stress relaxation in ordinary rubbers, which always has a very long-time tail (either power-law or even logarithmic). In `neat' transient networks (where the energy barrier for crosslink breaking has a well-defined value) the relaxation is strictly simple exponential, which allows us to determine the energy barriers. In `heterogeneous' transient networks where the energy barrier  for crosslink breaking is distributed over a wide range of values around a mean, the long-time stress relaxation follows a stretched-exponential law $\sim \exp[-(\beta t)^{0.2}]$. 
The key finding in the case of linear deformation ramp is the stress overshoot (yielding point) after which the network flows plastically. This yield point strongly depends on the applied strain rate. Finally, we examine the ability of transient networks to `self-heal', or recover the initial reference state when external forces are applied to keep it in that state for a sufficient length of relaxation time (which itself is a function of activation rate of crosslink breaking). 

Several approximations are made in this work to keep the transparency of the theory. We have omitted the non-affine movements of the system, which can be important when the chains between the crosslinks are short, or when the movement of entanglement is not negligible. The neo-Hookean model of rubber elasticity which we used is only strictly valid for small deformations, so a different elastic model should be used when dealing with large deformations when the chain inextensibility is tested. In spite of these limitations, we believe this work provide a clear and predictive picture of dynamics and relaxation in generic transient networks and offer insights for handling and processing such materials in practice.

\subsection*{Acknowledgements}
This work has been funded by the Theory of Condensed
Matter Critical Mass Grant from EPSRC (EP/J017639). We are grateful for the vitrimer samples donated by Prof. Yan Ji, Tsinghua University.


\providecommand{\latin}[1]{#1}
\providecommand*\mcitethebibliography{\thebibliography}
\csname @ifundefined\endcsname{endmcitethebibliography}
  {\let\endmcitethebibliography\endthebibliography}{}

\end{document}